\documentclass[a4paper]{PoS}

\def\siii{^3 \hskip -0.025in S _1}

\newcommand{\lsim}{\raisebox{-0.7ex}{$\stackrel{\textstyle <}{\sim}$ }}

\title{Magnetic properties of light nuclei from lattice QCD}

\ShortTitle{Magnetic properties of light nuclei from lattice QCD}

\author{\speaker{Martin J. Savage}\thanks{Presenting for the NPLQCD
    collaboration}\\
        Institute for Nuclear Theory,
        Box 351550,
        Seattle, WA 98195-1550, USA.\\
        E-mail: \email{mjs5@uw.edu}}

\abstract{
After a short review of Lattice QCD methodology and techniques,
I  summarize recent results of Lattice QCD calculations
of the interactions of nucleons and light nuclei 
with magnetic fields at pion masses of 805 MeV and 450 MeV. 
Interestingly, the magnetic moments
are found to be consistent with the experimental values when given in terms of natural nuclear magnetons.
The very low-energy cross section for $np\rightarrow d\gamma$ is calculated and found to agree 
with the experimental measurement.  
First calculations of the magnetic polarizabilities of light nuclei are presented, with
a large isovector polarizability observed for the nucleon at these heavier pion masses.
}

\FullConference{The 8th  International Workshop on Chiral Dynamics \\
29 June 2015 - 03 July 2015\\
Pisa, Italy}

\begin{document}

%%%%%%%%%%%%%%%%%%%%%%%%%%%%%%%%%%%%%%%%%%%%%%%%%%%%%%%%%%%%%%%%
%%     Introduction

\noindent
The electromagnetic interactions of nuclei have been used extensively to elucidate their structure and dynamics. In the early days of nuclear physics, 
the magnetic moments of the light nuclei helped to reveal that they behaved like a collection of {\it weakly} interacting nucleons that, to a very large degree, 
retained their identity, despite being bound together by the strong nuclear force. This feature, in part, led to the establishment of the nuclear shell model as a 
phenomenological tool with which to predict basic properties of nuclei throughout the periodic table. 
The strong nuclear force emerges from quantum chromodynamics
(QCD) as a by-product of confinement and chiral symmetry breaking. The fact that, at the physical values of the 
quark masses, nuclei are not simply collections of quarks and gluons, defined by their global quantum numbers, but have the structure of interacting 
protons and neutrons, remains to be understood at a deep level. 
In this presentation, I will show the results of recent Lattice QCD (LQCD) calculations of the magnetic moments~\cite{Beane:2014ora} 
and polarizabilities~\cite{Chang:2015qxa} of light nuclei at pion masses of
$m_\pi\sim 450~{\rm MeV}$ and $805~{\rm MeV}$, 
along with the first calculation of the radiative-capture cross section of $np\rightarrow d\gamma$~\cite{Beane:2015yha} extrapolated 
to nature and compared with experiment.

Lattice QCD is moving rapidly into a position to calculate the properties and interactions of the lightest nuclei at the physical values of the light-quark 
masses and with the inclusion of fully dynamical QED.   
The ultimate reason for such calculations is to refine the nuclear forces,
and enhance our  ability to reliably predict the properties and interactions of  nuclei throughout the Periodic Table,
beyond what is possible with experiment alone.
Currently, we are only now entering the ``verification stage'', where experimentally measured 
nuclear physics
observables that can be 
accessed by LQCD are 
demonstrated to be
reproduced within the uncertainties of the measurements and of the calculations.  
It will likely be another few years before calculations are of sufficient precision and accuracy 
to have verified LQCD as a reliable predictive tool for nuclear physics.
Until recently, it was only the scattering of two nucleons~\cite{Beane:2006mx,Ishii:2006ec,Murano:2013xxa,Beane:2013br} 
and the binding energies of the lightest nuclei and hypernuclei~\cite{Yamazaki:2009ua,Beane:2011iw,Beane:2012ey,Beane:2012vq,Yamazaki:2012hi,Yamazaki:2015asa} 
that were being pursued with LQCD because 
calculations of other properties, such as magnetic and quadruple moments, are not meaningful 
without clearly establishing a bound nucleus.  
Multi-meson systems have been considered previously as somewhat of a prelude to multinuclear systems, in particular the extraction of multi-body 
interactions~\cite{Beane:2007es,Detmold:2008fn,Detmold:2008yn,Detmold:2011kw}.

Conceptually, Lattice QCD calculations are a straightforward and brute-force evaluation of the QCD path integral for the observables of interest.
However, this is not quite QCD 
that is being calculated 
as the evaluation is  performed in Euclidean space with a finite lattice spacing providing the ultra-violet regulator and in a finite volume 
which modifies the infrared.
For a lattice spacing that is small compare with typical strong interaction length scales, the Symanzik action is used to systematically remove lattice spacing artifacts, 
usually through performing multiple calculations of the same quantities over a range of lattice spacings.
For a lattice volume that is large compared with the size of a compact state, such as a nucleus, effective field theory 
(EFT) techniques can be used to 
extrapolate to infinite volume,  
by performing calculations in multiple volumes.
Scattering states are a different matter, and the geometry of the finite-volume, along with the momentum of the system in the volume and the chosen boundary conditions, 
determines the nature of the eigenstates and their relation to Minkowski-space S-matrix elements.
One should view LQCD as technique to solve a low-energy EFT of quarks and gluons 
that is applicable for momenta that are much below the inverse lattice spacing of the lattice.
Typical lattice spacing used in present-day LQCD calculations are $a\lsim 0.1~{\rm fm}$, corresponding to a 
momentum scale of $\mu\sim 2~{\rm GeV}$.

\begin{figure}[!ht]
  \centering  
  \includegraphics[width=0.8\columnwidth]{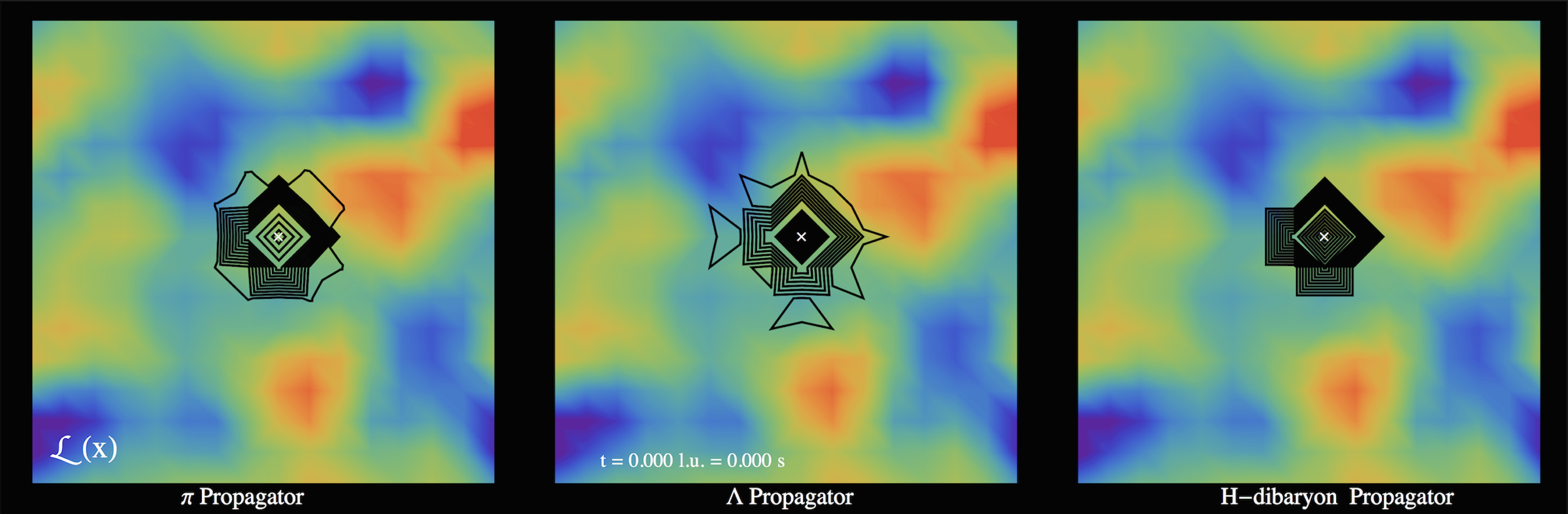}
  \includegraphics[width=0.8\columnwidth]{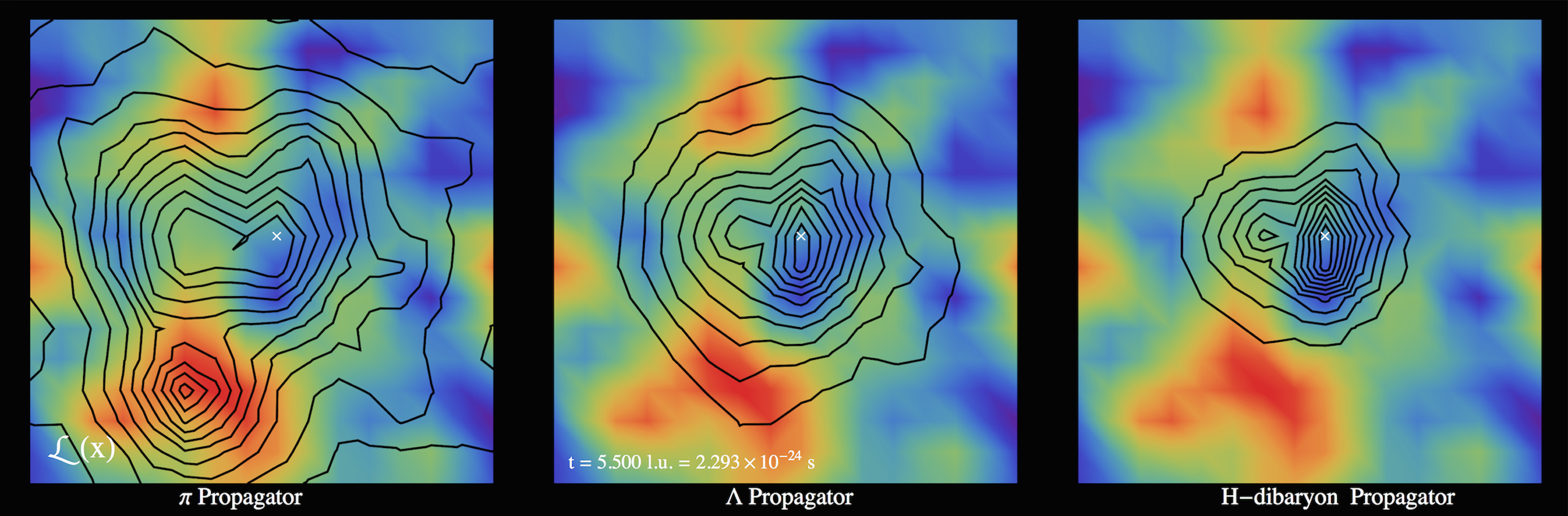}
  \includegraphics[width=0.8\columnwidth]{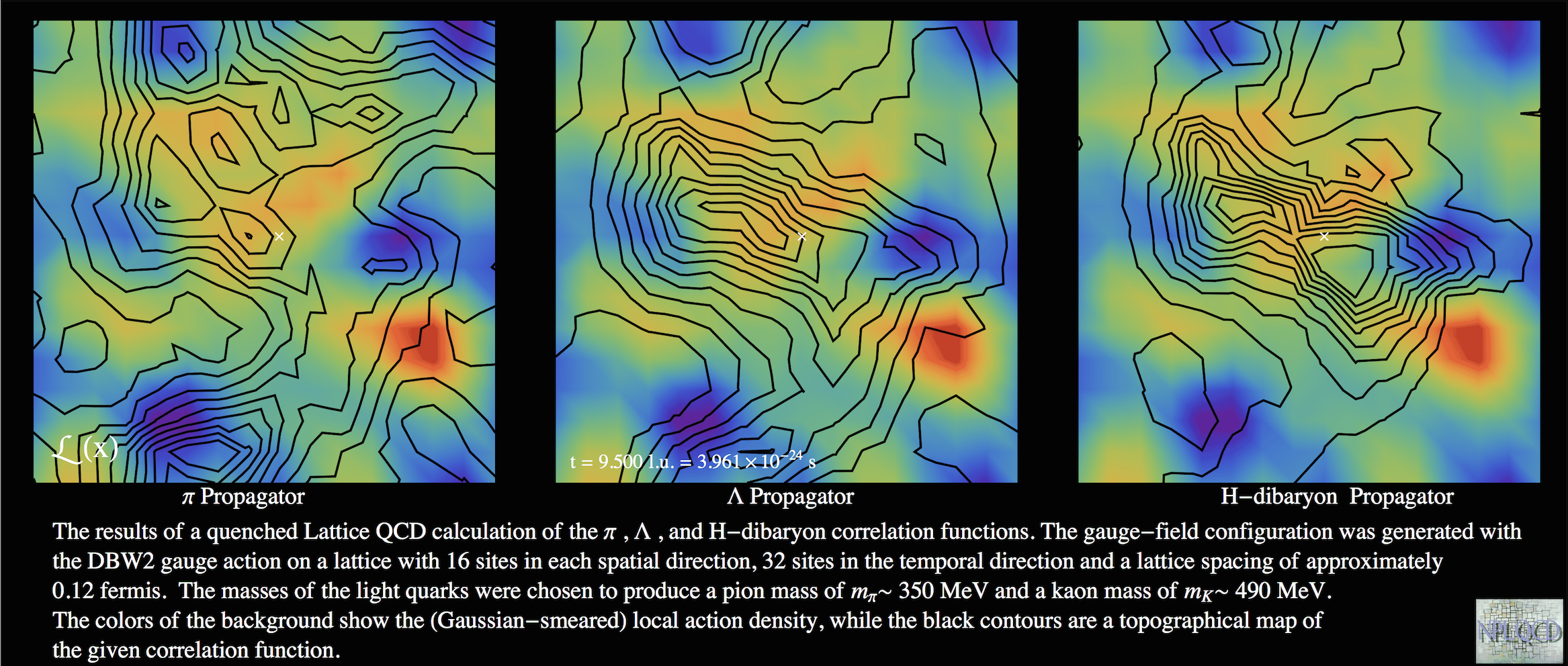}
\vskip -0.15in
  \caption{ The same spatial slice of the  correlation functions of the $\pi$, $\Lambda$ and $\Lambda\Lambda$ at three different times superposed on the action density of the gluon field.
  These ``toy''  calculations are quenched with a DBW2 gauge action and a pion mass of $m_\pi\sim 350~{\rm MeV}$.
     }
\label{fig:correlationfunctionsgaugefield}
\end{figure}

One of the important features of LQCD calculations of systems involving baryons (compared to mesons) 
is the signal-to-noise degradation that is encountered in the correlation functions.
At large times, the signal-to-noise ratio in a nucleon correlation function scales as $e^{-(M_N-3 m_\pi/2)t}$, but degrades somewhat less severely at 
intermediate times by judicious choices of interpolating operators.  
Insight can be gained into the behavior of correlation functions from the results of calculations, such as those shown in Fig.~\ref{fig:correlationfunctionsgaugefield}.
The fluctuations in the gauge field clearly impact the evolution of the hadron correlation functions in highly non-trivial ways, as one expects.  
These interactions introduce significant non-Gaussianity into the distribution of correlations functions on any particular time-slice.

NPLQCD has performed a comprehensive study of s-shell nuclei and hypernuclei,
and of nucleon-nucleon scattering, at the flavor-SU(3) symmetric point at  pion mass of 
$m_\pi\sim 805~{\rm MeV}$, where the strange-quark mass is tuned to its value in nature.
Calculations were performed in multiple lattice volumes, 
parametrically reducing the residual finite-volume effects.
However, due to computational resource limitations, calculations were performed at one lattice spacing only, 
but one that is estimated to be small enough to have only small effects on the values of 
calculated binding energies and scattering parameters.
\begin{figure}[!ht]
  \centering  
  \includegraphics[width=0.8\columnwidth]{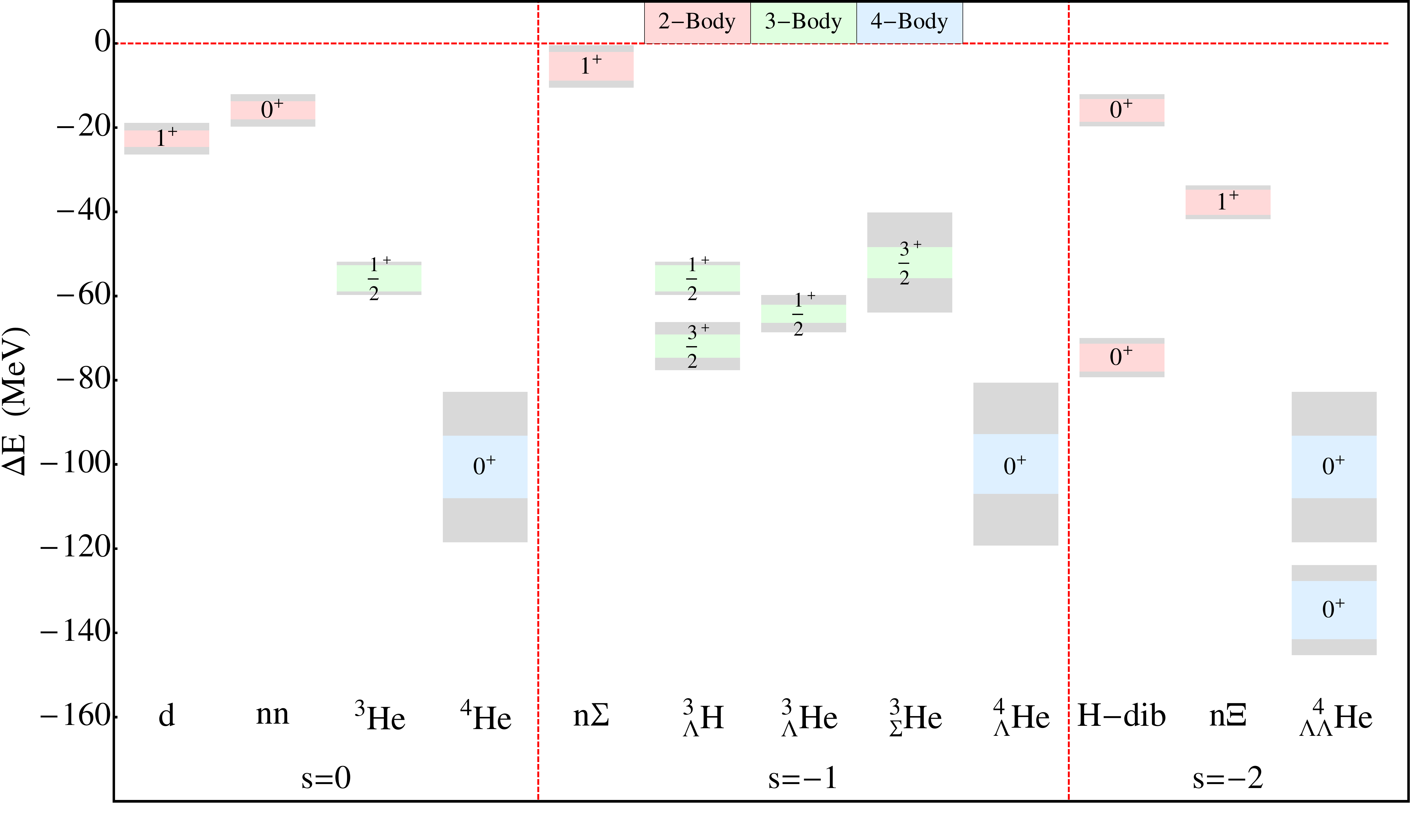}
 \vskip -0.15in
  \caption{ The spectrum of the s-shell nuclei and hypernuclei  obtained 
by the NPLQCD collaboration~\protect\cite{Beane:2012vq}
  at the flavor-SU(3) symmetric point with a pion mass of 
$m_\pi\sim 805~{\rm MeV}$.
     }
\label{fig:SU3Bindings}
\end{figure}
The calculated binding energies are shown in Fig.~\ref{fig:SU3Bindings}.
Ground states, and in some cases excited states, are well resolved for nuclei and hypernuclei with atomic number $A=1,2,3$ and $4$.
Isolating the ground states is necessary for further studies of their  properties, such as magnetic moments and polarizabilities.
\begin{figure}[!ht]
  \centering  
  \includegraphics[width=0.475\columnwidth]{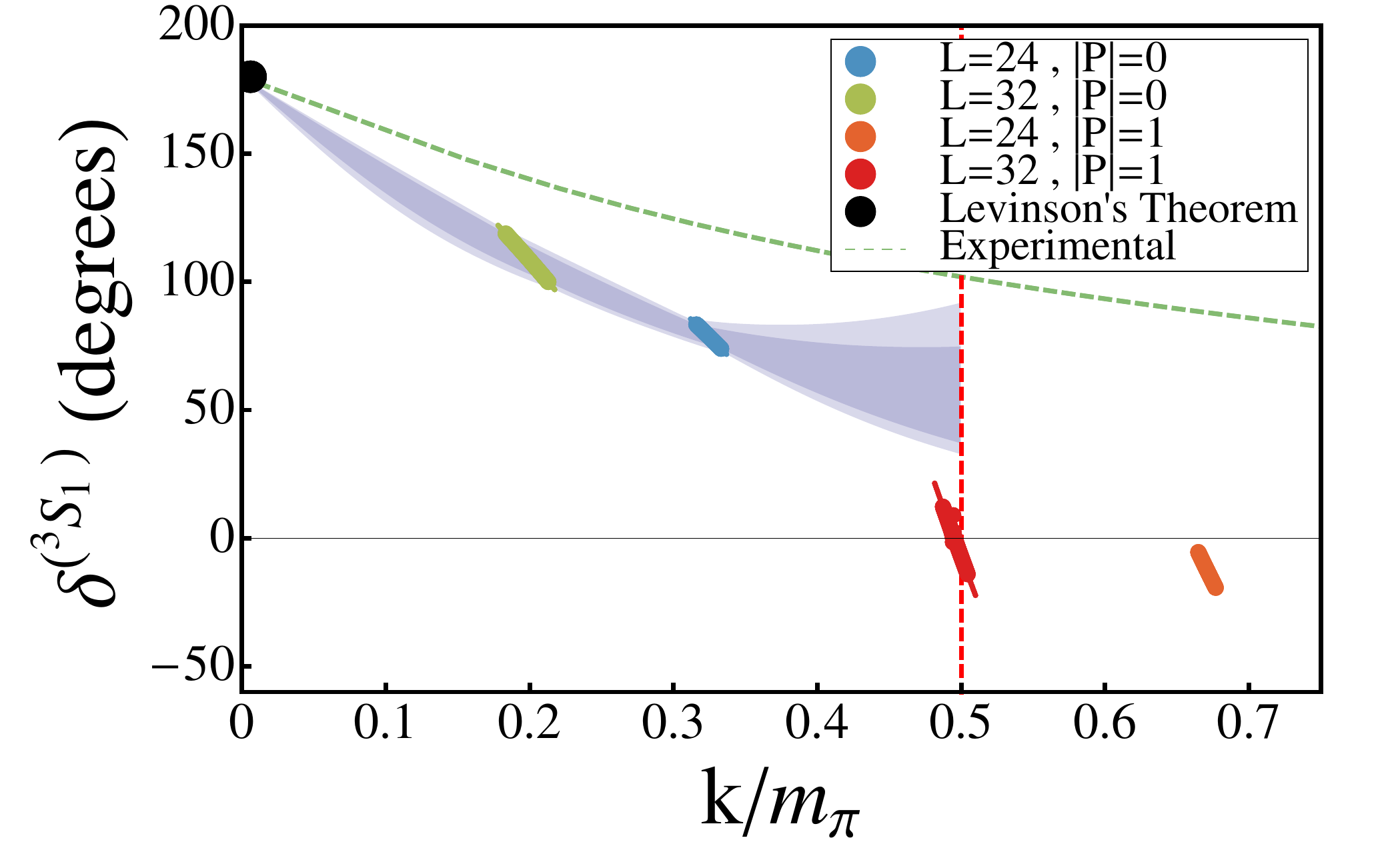}
  \includegraphics[width=0.475\columnwidth]{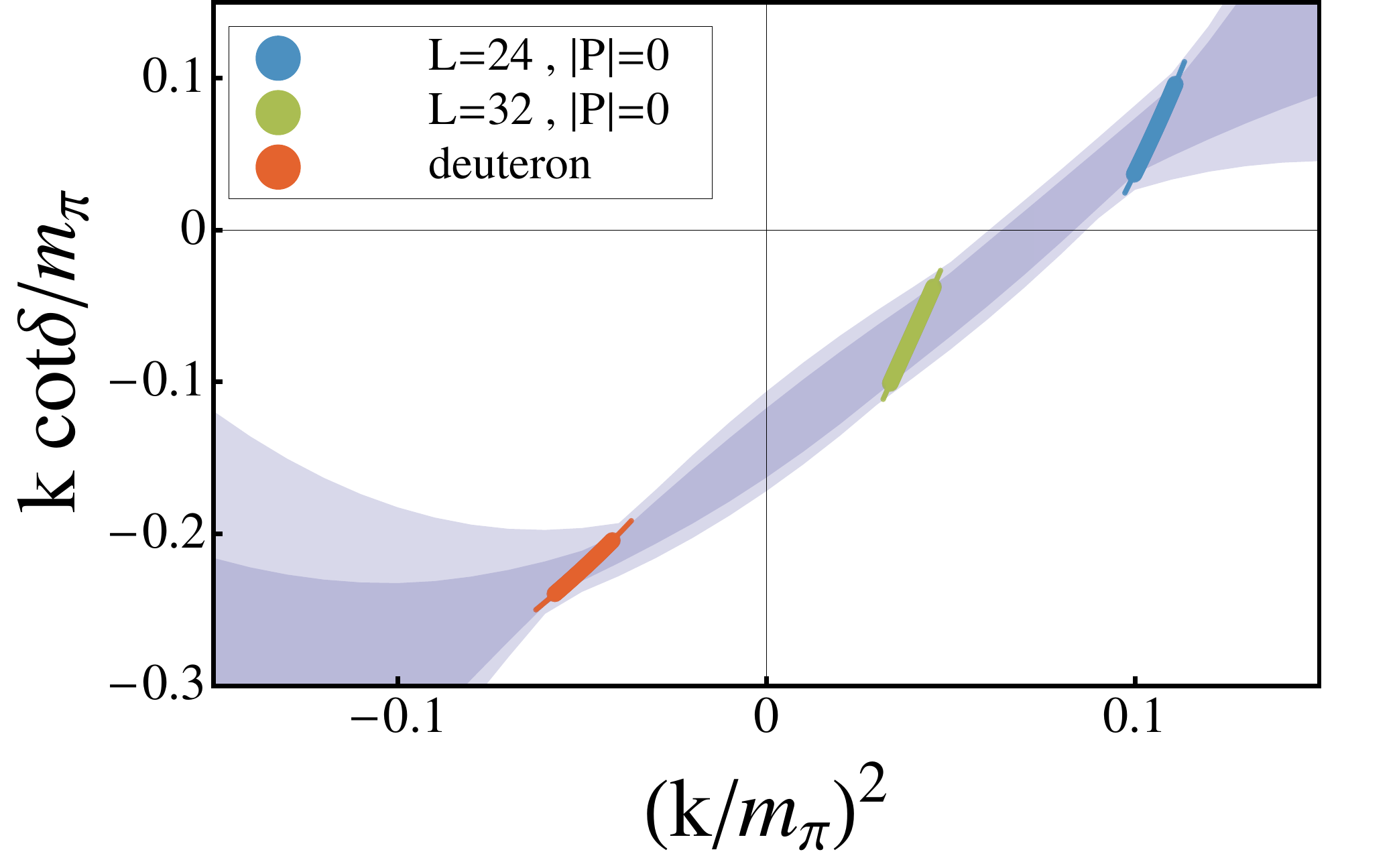}
 \vskip -0.15in
  \caption{ Nucleon-nucleon scattering phase shifts (left panel) and 
  $k\cot\delta$ (right panel) in the $\siii$-channel 
  at the flavor-SU(3) symmetric point with a pion mass of 
$m_\pi\sim 805~{\rm MeV}$~\protect\cite{Beane:2013br}.
     }
\label{fig:NNSU3}
\end{figure}
The scattering phase shift in the $\siii$ channel,
shown in Fig.~\ref{fig:NNSU3}, 
is extracted from the location of higher-lying states in the spectrum using Luscher's quantization 
conditions (and assuming higher partial-wave phase shifts are small).
Also shown in  Fig.~\ref{fig:NNSU3} is the extracted values of $k\cot\delta$ that uniquely defines the nucleon-nucleon scattering amplitude below the start of the t-channel cut.
The precision of the calculation allows for a determination of both the scattering length and effective range.  The scattering length is found to be twice as large as the effective range, suggesting that the deuteron remains a somewhat large object over a significant range of quark masses, and is not finely-tuned as had been thought previously. 
This intriguing feature requires further investigation.

\begin{figure}[!ht]
  \centering  
  \includegraphics[width=0.8\columnwidth]{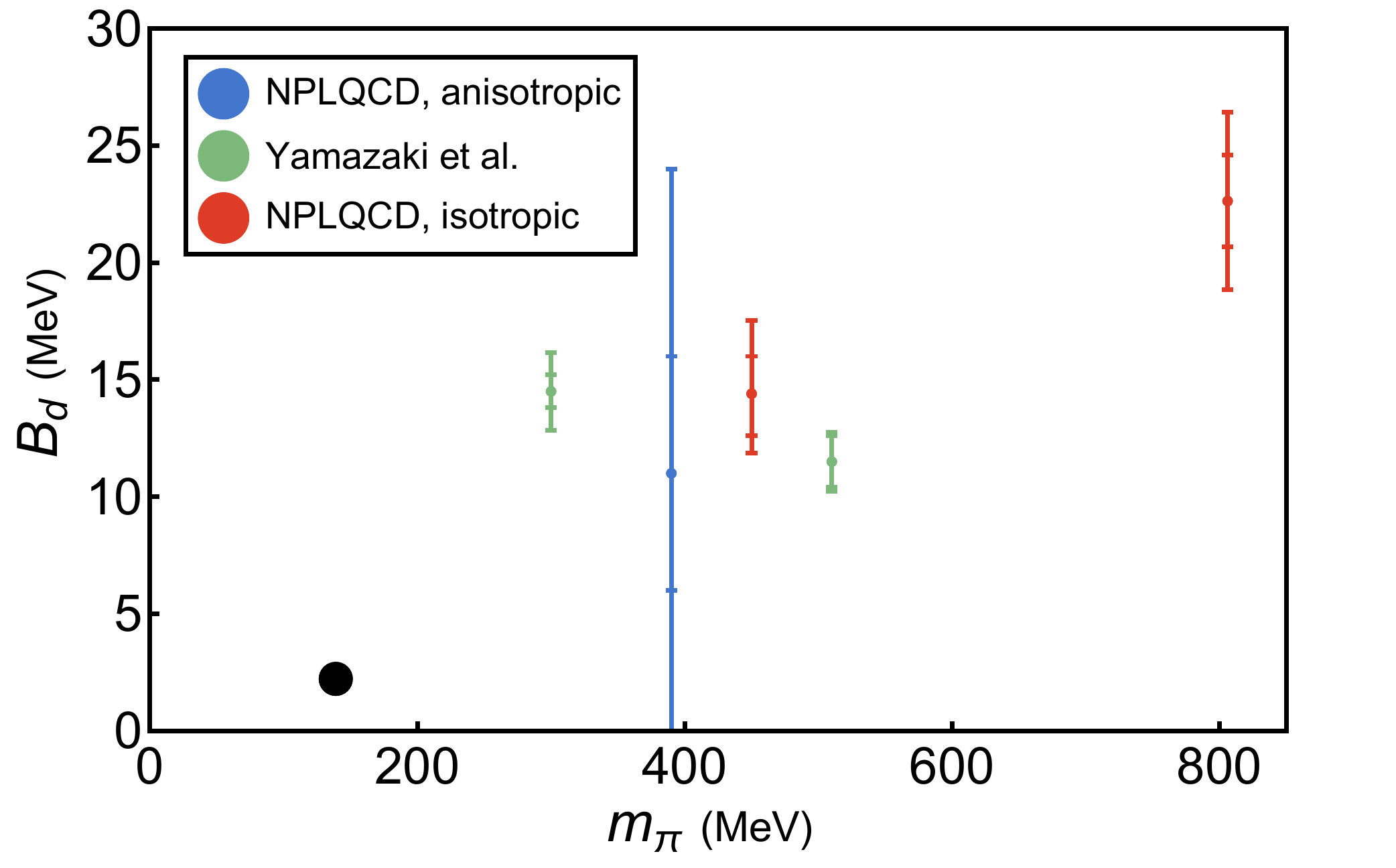}
 \vskip -0.15in
  \caption{ Summary of fully-dynamical Lattice QCD calculations of the deuteron binding energy as a function 
  of the pion mass~\cite{Beane:2011iw,Beane:2012vq,Yamazaki:2012hi,Yamazaki:2015asa,Detmold:2015:450MeV}.  
  }
\label{fig:DeutAll}
\end{figure}

When the results from NPLQCD~\cite{Beane:2011iw,Beane:2012vq,Detmold:2015:450MeV} are combined with those from 
Yamazaki {\it et al}.~~\cite{Yamazaki:2012hi,Yamazaki:2015asa}, 
a pattern  begins to emerge as to how the deuteron binding energy varies with the light-quark masses, as shown in 
Fig.~\ref{fig:DeutAll}~\footnote{
It should be mentioned that the calculations of the HAL QCD collaboration do not find bound nuclei at these heavier pion masses, 
e.g. Ref.~\cite{Aoki:2012tk}.  
However,   they have not directly calculate 
bindings, but have used a uncontrolled truncation of a non-local lattice correlation function
to arrive at their conclusion(s).  The apparent disagreement is likely due to the limitations of this hadronic modeling.  
For more discussion on this technique, see Refs.~ \protect\cite{Beane:2008dv,Beane:2010em} .
}.
It appears that nothing dramatic is going to happen to the deuteron binding as the light-quark masses 
are reduced to their physical values,
 and that the system evolves smoothly to the experimentally measured binding energy.
Calculations at a pion mass of $m_\pi\sim 300~{\rm MeV}$ are ongoing, and we expect that calculations closer to  $m_\pi\sim 140~{\rm MeV}$ will be performed soon, see Ref.~\cite{{YamaPHYS2015}}.

Having identified ground state plateaus of the lightest nuclei, we are now in a position to ask questions about their structure.  
Historically, the magnetic moments of nuclei, and magnetic transitions between states, have revealed much about their structure.
In particular, that they are  close to the values predicted by the naive nuclear shell model.
Exploring the range of validity of the nuclear shell model as a function of the light-quark masses - a dimension that is not accessible experimentally - 
may reveal further secrets 
about the nature of the strong interactions and the structure of nuclei.  In fact, it does.
There are a couple of techniques with which to determine the magnetic moments of hadrons with LQCD calculations,
and  
NPLQCD chose to use background magnetic fields in its studies.  
\begin{figure}[!ht]
  \centering  
  \includegraphics[width=0.8\columnwidth]{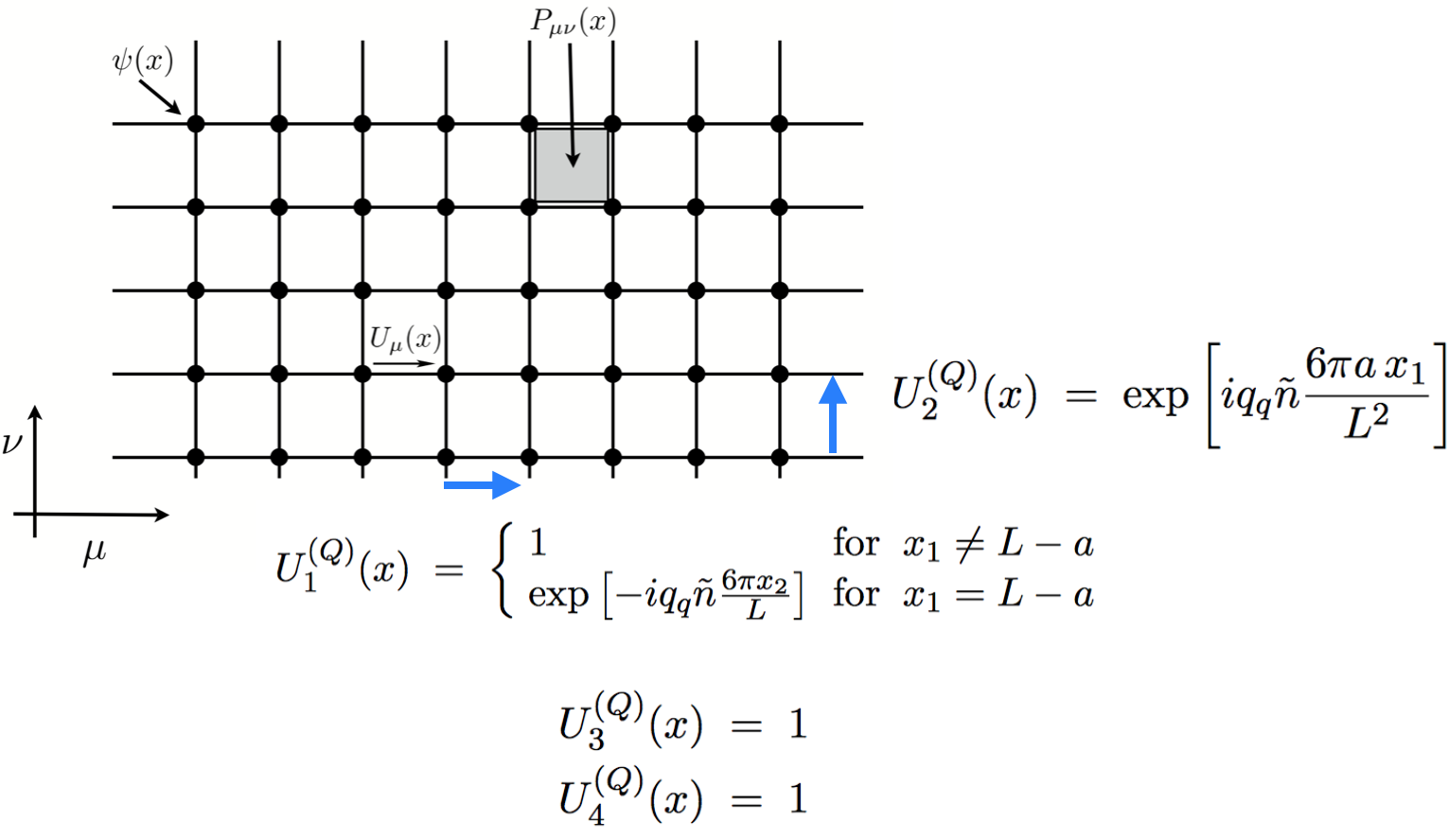}
 \vskip -0.15in
  \caption{ A $U(1)$ gauge field that generates a uniform and time-independent magnetic field 
  $e{\bf B} = {6\pi \tilde n\over L^2} \hat {\bf e_z}$.
     }
\label{fig:bkgdlattice}
\end{figure}
A $U(1)$ gauge field is constructed to produce a uniform and time-independent magnetic field (along the z-direction for simplicity).
It is required to be quantized 
(in units of $\tilde n$), 
as shown in Fig.~\ref{fig:bkgdlattice}.
Each configuration of QCD gauge fields in a given ensemble is post multiplied by this $U(1)$ gauge field, which subjects the valence 
quarks to the magnetic field, but  not the sea quarks.  
Having the sea quarks also interact with the  magnetic field would require it to be 
present during the generation of the ensemble, which is currently beyond our computational resources.
Therefore, the results of such calculations are generally not predictions of QCD, due to the absence of electromagnetic coupling to the sea quarks, 
however there are certain quantities that are predictions of QCD.  For instance, at the flavor-SU(3) symmetric point, the electric charge matrix is traceless, 
and as such there is no coupling of one insertion of the magnetic field to the sea quarks. 
Therefore, the magnetic moments of the hadrons and nuclei calculated with such a post-multiplication by the background field are equivalent to those generated with fully-dynamical QED up to terms that are higher order in $\alpha_e$.
In contrast, while the isovector polarizabilities, corresponding to two insertions of the magnetic field, are equivalent to a full calculation, 
the isoscalar polarizabilities are not due to loop diagrams involving two photons coupling to one sea-quark loop.   
The complexity is more significant away from the SU(3) point as loops involving one photon no longer vanish by the 
tracelessness of the charge matrix due to SU(3) breaking effects.

A  complication of LQCD calculations in the presence of a background magnetic field is that the energy eigenstates 
of charged states are no 
longer momentum eigenstates, but are Landau levels in the plane transverse to the field and momentum eigenstates along the direction of the field.
In dimensionless units, the energy difference of the ground state of a nucleus, $h$,  
of spin $j$ and projection $j_z$ along a magnetic field along the z-direction
(${\bf B}= B {\bf e_z}$ and $P_\parallel=0$) is 
\begin{eqnarray}
a\ \delta E_{h;j_z} 
&=&  \sqrt{a^2 M_h^2 + (2n_L+1)Q_h a^2|e\ {B}|} - a M_h 
- \frac{2 e}{ a M_N} \hat \mu_h j_z a^2 |e\ B| 
\nonumber \\
&& 
\qquad 
- \frac{2\pi}{a^3 M_N^2(M_\Delta-M_N)} \left[\hat \beta_h 
+ \hat \beta_h^{(2)}  ( j_z^2 - {1\over 3} j(j+1) )  \right]\left(a^2 |e\ B|\right)^2
\nonumber \\
&& 
\qquad +j_z \hat \gamma_h \left(a^2 |e\ B|\right)^3
+ \hat \delta_h \left(a^2 |e\ B|\right)^4
\ +\ ....
\ \ \ ,
\label{eq:Bdependlatt}
\end{eqnarray}
where
$n_L$ is the integer associated with the Landau level,
$\hat \mu_h = \mu_h {M_N\over 2e}$ is the dimensionless magnetic moment,
both $\hat\gamma_h $ and $\hat\delta_h$ are terms higher order in the expansion
(but which are present to stabilize fits to the magnetic moments and polarizabilities).
The dimensionless polarizability and tensor polarizability are
\begin{eqnarray}
&&
\hat\beta_h = \frac{M_N^2(M_\Delta-M_N)}{e^2} \beta_h^{(M0)}  ,\ \ 
\hat\beta_h^{(2)} =  \frac{M_N^2(M_\Delta-M_N)}{e^2} \beta_h^{(M2)} \,,
\label{eq:imlessvardef}
\end{eqnarray}
and $a^2|e\ B|=\frac{6\pi \tilde n}{L^2}$ is the dimensionless field strength.

\begin{figure}[!ht]
  \centering  
  \includegraphics[width=0.6\columnwidth]{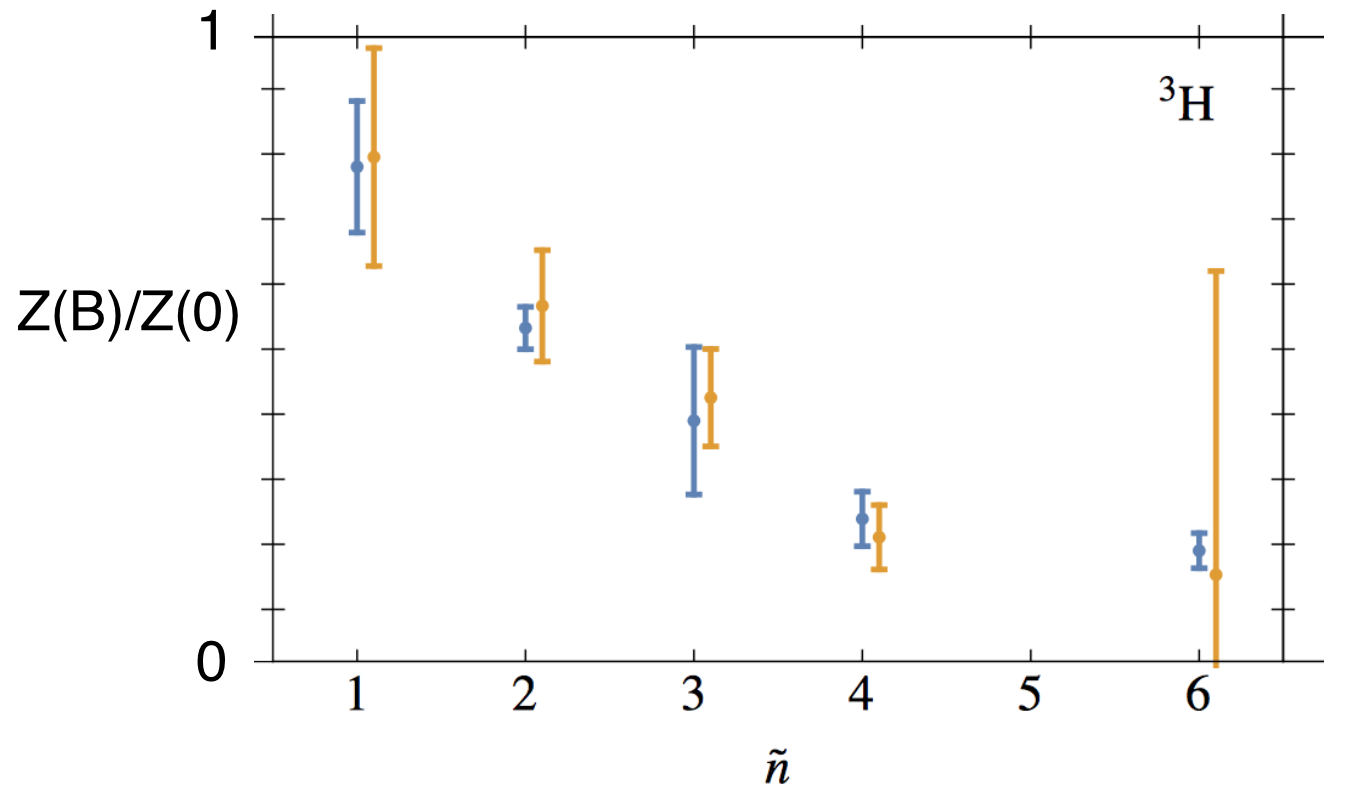}
 \vskip -0.15in
  \caption{ The ratio of  overlaps of the hadronic blocks onto the triton ground state as a function of  magnetic 
  field~\cite{Chang:2015qxa}. 
       }
\label{fig:ZBZzero}
\end{figure}

In the absence of a magnetic field, nuclear correlation functions are constructed from ``hadronic blocks''.  
These are three light-quark propagators  contracted at the sink with the quantum numbers of a 
proton or neutron, and then momentum projected.
This method of constructing nuclear correlation functions has proven to be efficient and successful for  light nuclei.
However, in the presence of a magnetic field, the Landau levels 
associated with a proton are different from a nucleus, say the triton, as both the charge and the mass 
determine the orbit. 
 Therefore, it was anticipated that the hadronic blocks might not be optimal for the production of nuclear correlaton functions 
 in a magnetic field.
 This was observed numerically, as shown in the ratio $Z(B)/Z(0)$ for the triton ground state in Fig.~\ref{fig:ZBZzero}.
The calculations were most precise for the neutron and di-neutron systems for which there are no Landau levels.

\begin{figure}[!ht]
  \centering  
  \includegraphics[width=0.48\columnwidth]{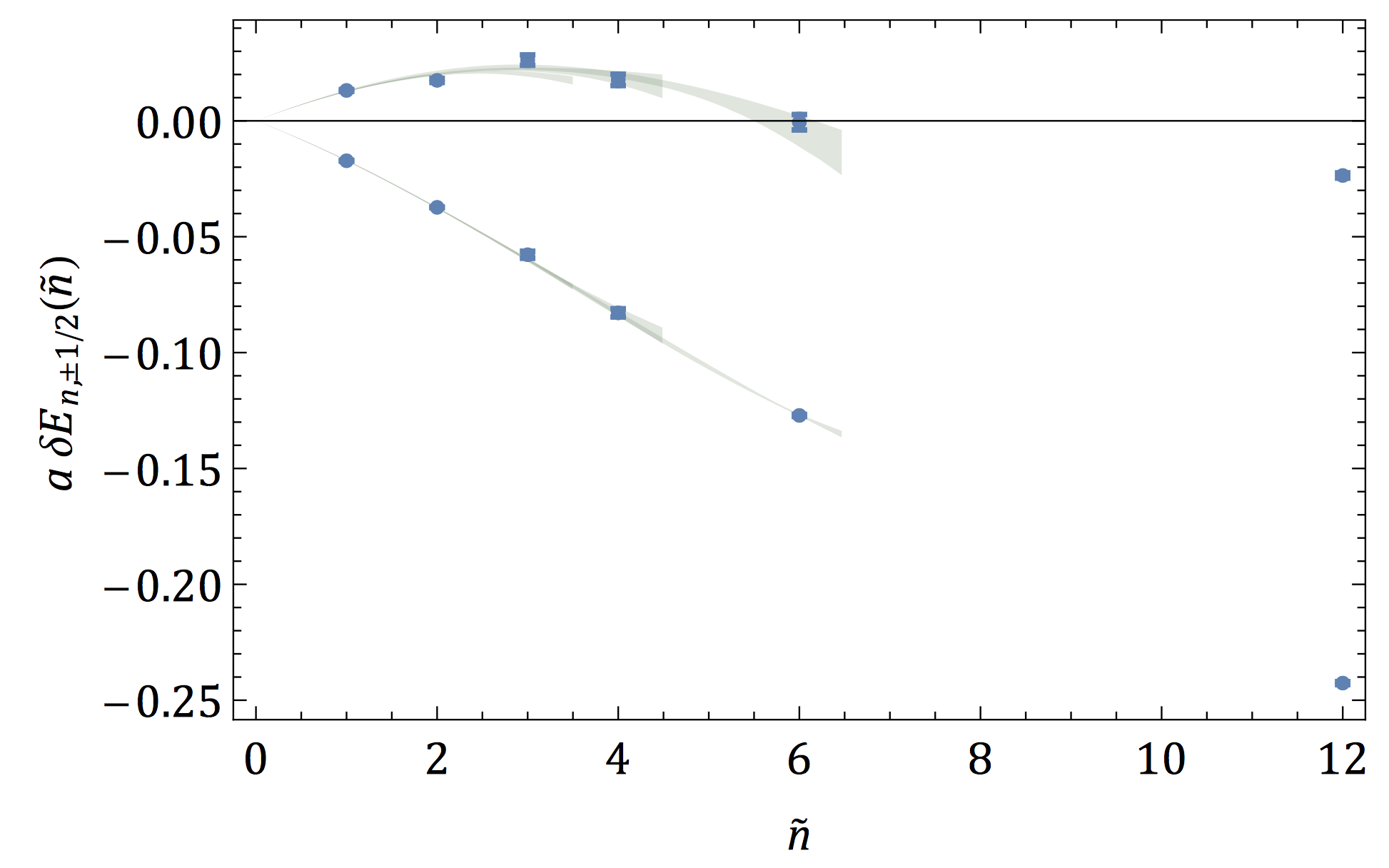}
  \includegraphics[width=0.48\columnwidth]{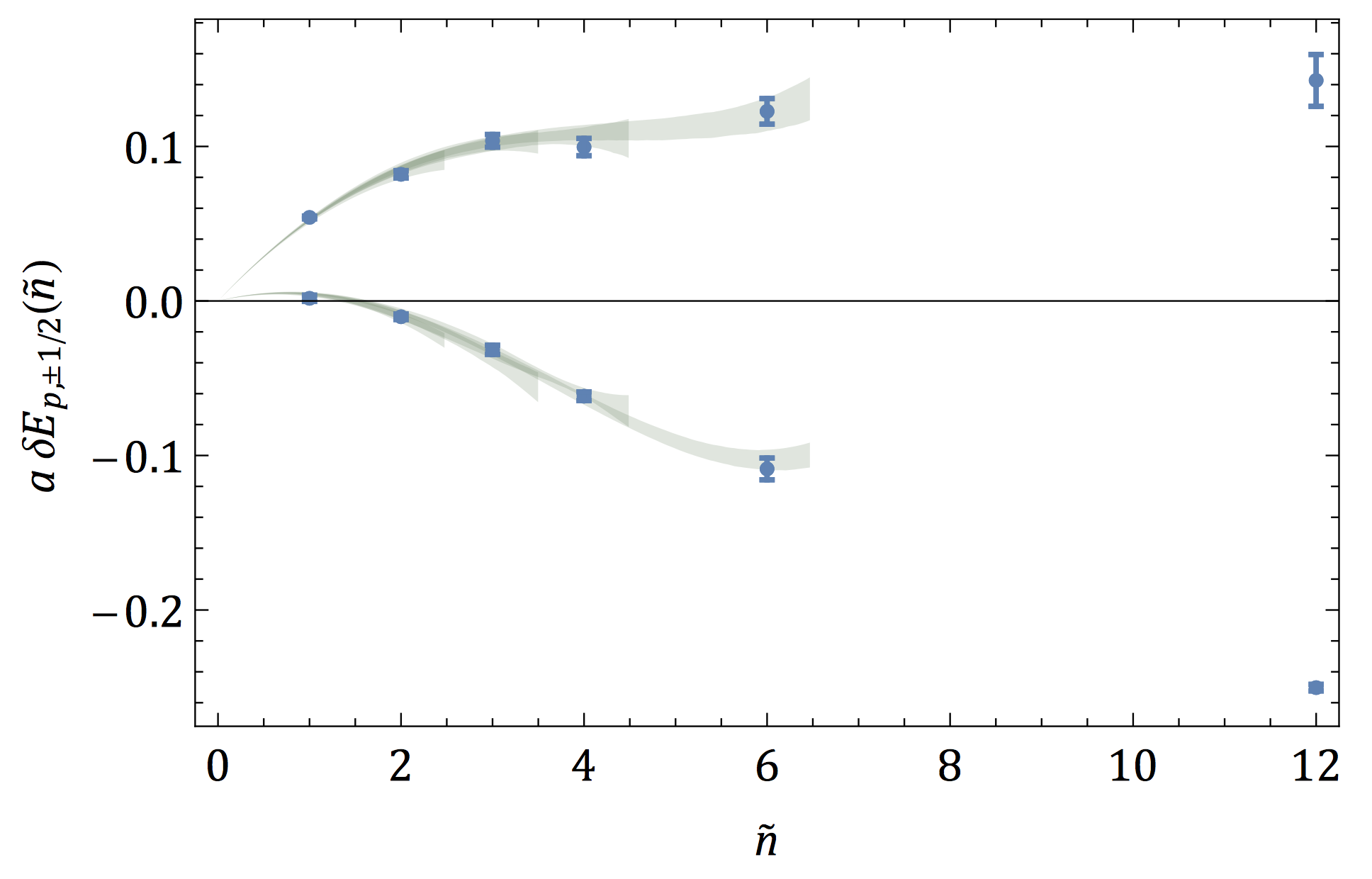}
 \vskip -0.15in
  \caption{
  The energy shifts induced in each spin state of the neutron (left panel) and proton (right panel) by the 
  magnetic field~\cite{Chang:2015qxa}.
       }
\label{fig:nstates}
\end{figure}

For the simplest case of the neutron, the negatively-shifted spin-down state depends essentially linearly on the 
magnetic-field strength out to very high magnetic fields, as shown in Fig.~\ref{fig:nstates}.
For the fields used in this work, $|e{\bf B}|\sim 0.05 \tilde n~{\rm GeV}^2$, and therefore $\tilde n=1$ corresponds to $|{\bf B}|\sim 10^{19}~{\rm Gauss}$.
In physical units, the energy shift in the $\tilde n=12$ magnetic field is $\Delta E\sim 400~{\rm MeV}$.
Therefore, we see the lower-lying state depend linearly upon the magnetic field out to about $\sim 10^{20}~{\rm Gauss}$.
The upper level, on the other hand, shows extensive curvature setting in at quite low fields (on this scale), 
and has similarities with what one observes in avoided crossing in two-state systems (with the other state not observed in this case).
It is clear from this behavior that the magnetic polarizability of the neutron is determined entirely by the curvature of the upper state, and receives little or no contribution from the lower state.
An interesting question to ask, that obviously requires significantly finer lattices, is to how the states behave in 
the limit of asymptotically large magnetic fields, where one expects only the behavior of nearly massless charged quarks to dominate the spectrum.
In particular, when does the expected curvature set in for the lower state?
The behavior of the proton states is more complex due to the Landau-level structure.  However, the system can be analyzed and the magnetic moment and polarizability extracted in a similar way to that of the neutron. 
The uncertainties are larger as the correlation functions are of somewhat poorer quality due to smaller overlap factors.
Further, the landau level associated with the plateau on the correlation function is unknown and must be determined by fitting~\cite{Chang:2015qxa}.
\begin{figure}[!ht]
  \centering  
  \includegraphics[width=0.8\columnwidth]{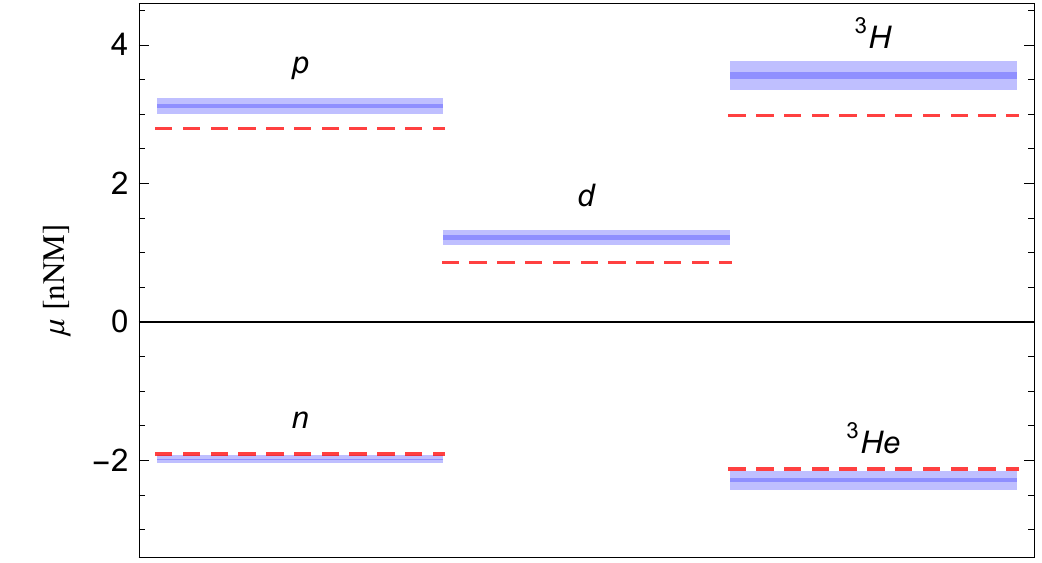}
 \vskip -0.15in
  \caption{
  The results of the Lattice QCD calculations of nuclear magnetic moments~\protect\cite{Beane:2014ora} are shown as the blue bands, 
  while the corresponding experimental values are shown by the red dashed lines.
  Natural nuclear magnets have been used, in which the nuclear magneton is defined in terms of the nucleon mass 
  at the given value of the light-quark masses.
         }
\label{fig:Amagmoms}
\end{figure}
A summary of the magnetic moments of the nucleon and lightest nuclei is shown in Fig.~\ref{fig:Amagmoms}.
It is remarkable that, when given in terms of natural Nuclear Magnetons (defined with the nucleon mass at the given light-quark masses), 
the magnetic moments of the light nuclei at a pion mass of $m_\pi\sim 805~{\rm MeV}$ are very close to  their values at the physical light-quark masses.
This implies that essentially all of the light-quark mass dependence of the magnetic moments is determined by the mass of the nucleon, and that a non-relativistic quark model type scenario is providing the dominant contribution - one in which a naive weighted sum of quark-model quark spins is dominant.  This could entirely be a consequence of the large-$N_c$ limit of QCD or it might be something more.
Another, and perhaps more remarkable,  feature is that the magnetic moments of the nuclei are essentially given by the sum of the nucleon magnetic moments in a naive nuclear shell-model configuration.   For these nuclei, this is somewhat trivial compared to larger nuclei, but nonetheless the two neutrons in the triton are largely in a spin-zero configuration with the magnetic momentum being essentially that of the proton.  The deviations observed in nature from the naive shell-model prediction are in agreement, within uncertainties, with the result of the LQCD calculation at the heavier pion mass.
This leads one to observe that nuclei behave as collections of {\it weakly} interacting nucleons over a large range of light-quark masses, and that the 
phenomenological nuclear shell-model is not limited in applicability to the physical point, but is somewhat of a generic feature of QCD for arbitrary light-quark masses.  It will be interesting to learn if there are values of the quark masses for which nuclei  collapse into a strongly interacting configuration of quarks and gluons rather than of weakly nucleons with a  hierarchy of multi-nucleon forces.
\begin{figure}[!ht]
  \centering  
  \includegraphics[width=0.6\columnwidth]{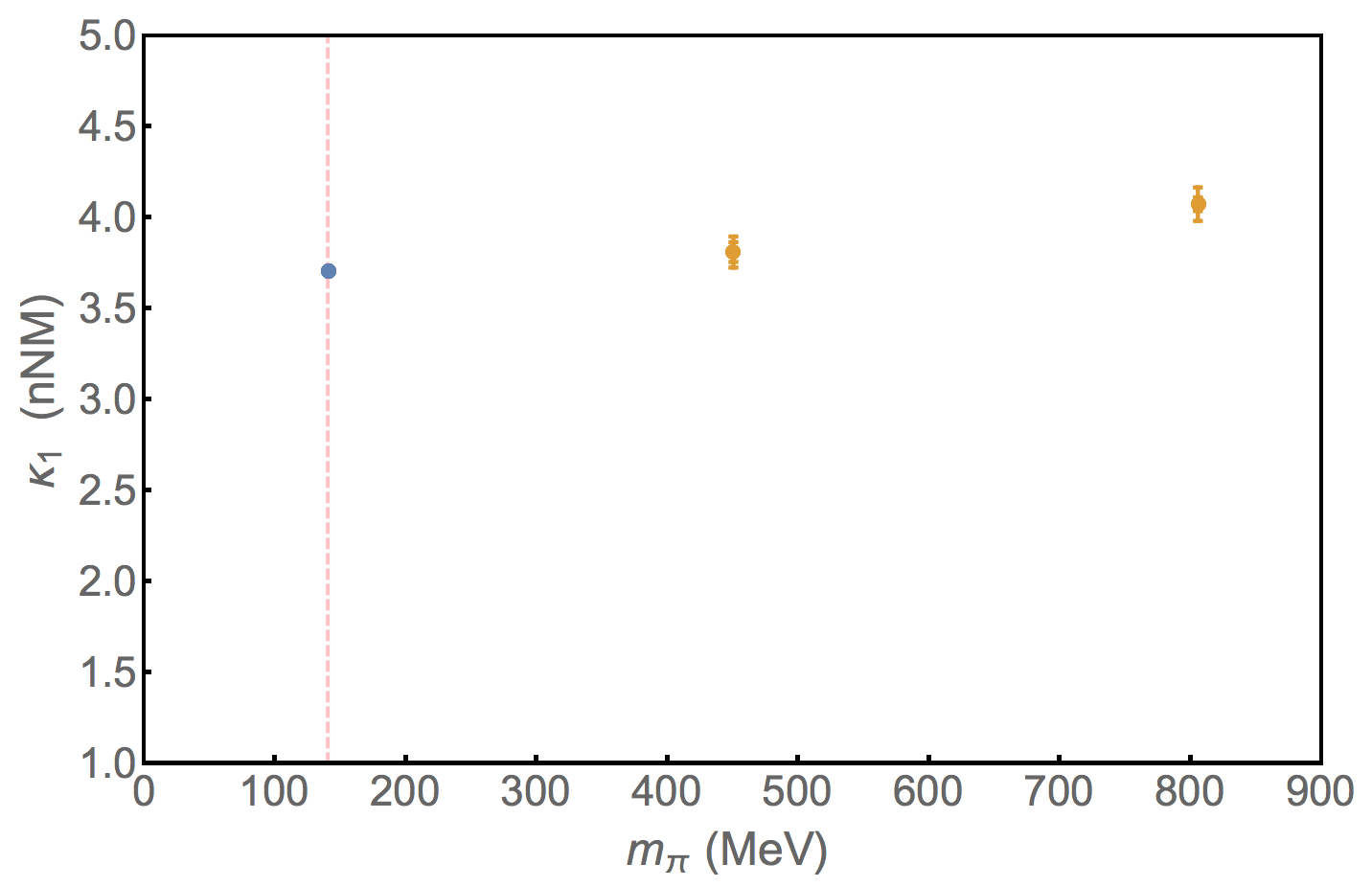}
 \vskip -0.15in
  \caption{
  The isovector anomalous magnetic moment of the nucleon [from the NPLQCD collaboration].
         }
\label{fig:kappaN}
\end{figure}
The magnetic moments have also been calculated at a pion mass of $m_\pi\sim 450~{\rm MeV}$.  
The same behavior is observed, as demonstrated in the 
isovector anomalous magnetic moment of the nucleon shown in Fig.~\ref{fig:kappaN}.

A key reaction in Big Bang  nucleosynthesis is the radiative capture process $np\rightarrow d\gamma$, which is the primary mechanism to 
first produce deuterium.  It has been long known that a naive calculation of the cross section for this process using only nucleon-nucleon  scattering 
S-matrix elements provides only $\sim 90\%$ of the experimentally measured cross section at low energies.  
It has also been long recognized that this defficiency is due to the relatively large role played by meson-exchange currents (MECs), or alternatively from a modern perspective the role of correlated short-distance two-nucleon interactions with the electromagnetic field.  Once the contributions from these interactions are included, the cross section comes into agreement with 
experiment.~\footnote{
In low-energy EFTs, 
there is an M1 local operator with an coefficient that cannot be determined from scattering alone~\protect\cite{Kaplan:1998sz,Chen:1999tn}, 
and hence agreement at one kinematic point is somewhat trivial.  
However, after precisely determining this counter term, there is good agreement in the M1 amplitude over a range of 
energies~ \protect\cite{Chen:1999bg,Rupak:1999rk,Tornow:2003ze} that would otherwise have 
been in conflict.}
With the same set of LQCD calculations that provided the nuclear magnetic moments (at both $m_\pi=450~{\rm MeV}$ and $805~{\rm MeV}$), the splitting between the $j_z=0$ neutron-proton energy levels in the background magnetic field allowed for a determination of the  correlated short-distance two-nucleon interactions with the magnetic field, and hence for 
the first QCD prediction of $np\rightarrow d\gamma$~\cite{Beane:2015yha}.  
At the heavier pion mass, this enabled  a prediction of the cross section without any experimental input - truly 
a QCD prediction, while at the physical pion mass, the scattering parameters were used, in part,  to predict the complete cross section.
Using the physical scattering parameters and the LQCD calculations of the correlated short-distance two-nucleon interactions with the magnetic field, a cross section of $\sigma^{\rm lqcd}=332.4^{+54.}_{-4.7}~{\rm mb}$ is calculated at a neutron incident speed of $v=2200~{\rm m/s}$, which is to be compared with the experimental value of  $\sigma^{\rm exit}=334.2\pm 0.5~{\rm mb}$.

The curvature of the energy of the nucleon or nucleus, after removing the contribution from the Landau level, provides a determination of its magnetic polarizability.
\begin{figure}[!ht]
  \centering  
  \includegraphics[width=0.48\columnwidth]{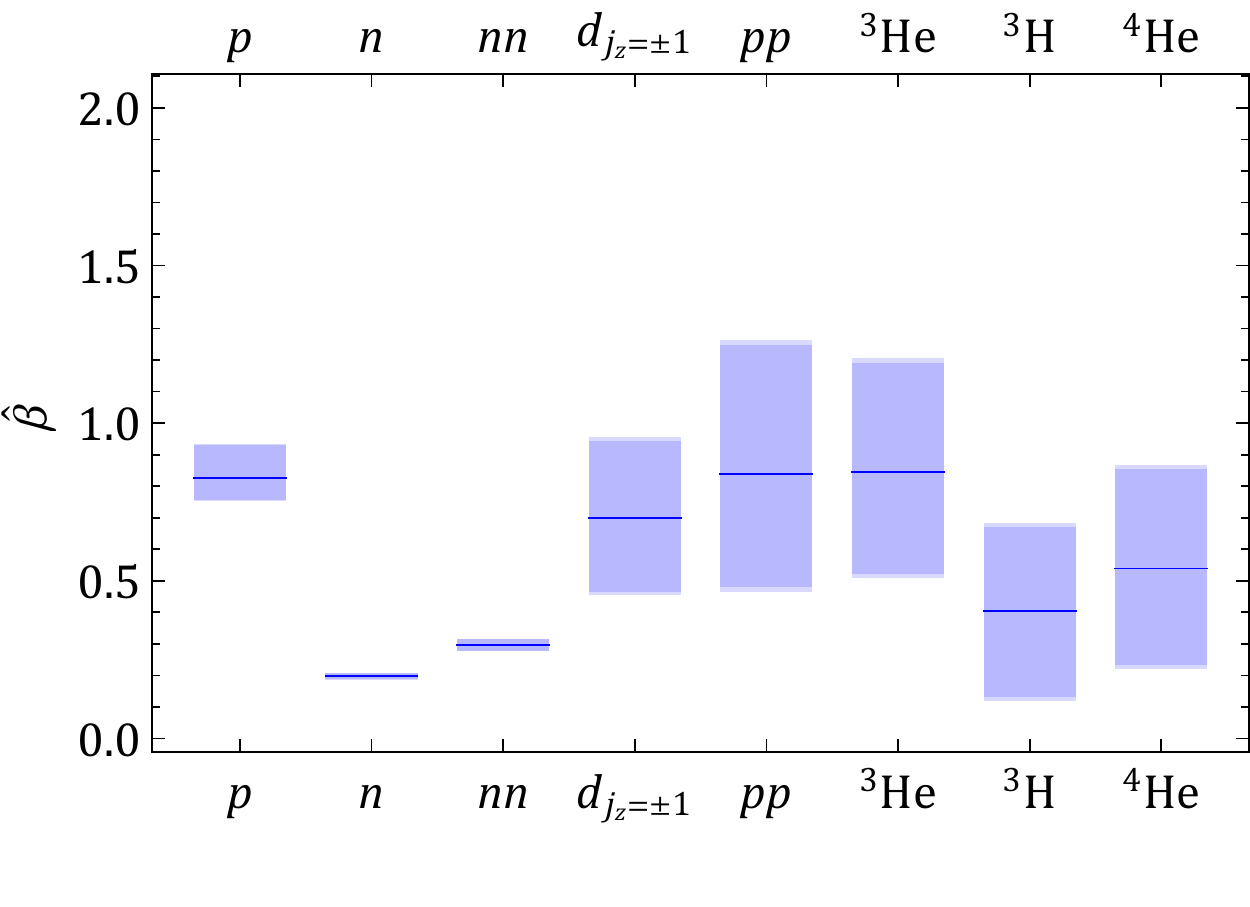}
  \includegraphics[width=0.48\columnwidth]{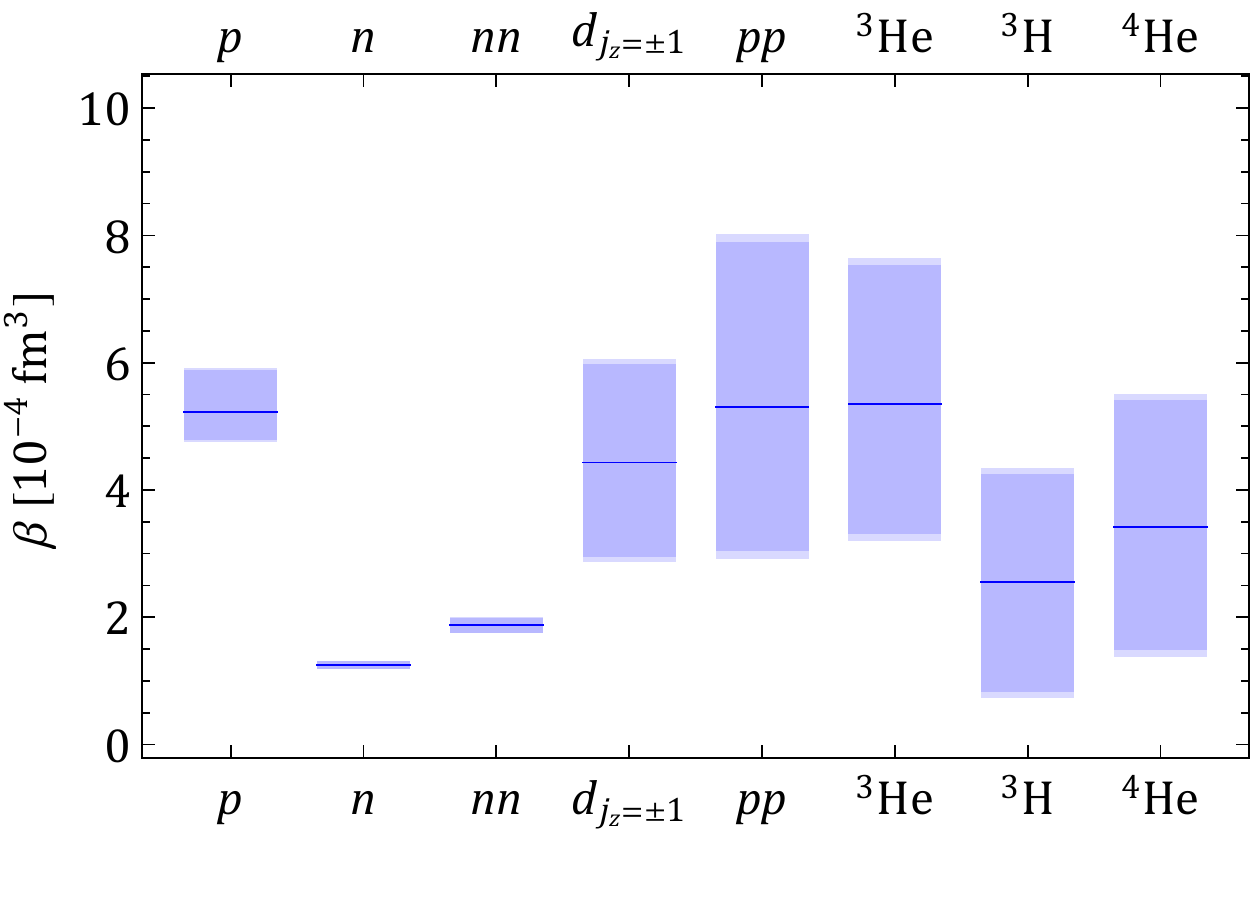}
 \vskip -0.15in
  \caption{
  The magnetic polarizabilities of the lightest nuclei at a pion mass of $m_\pi\sim 805~{\rm MeV}$~\cite{Chang:2015qxa}.  
  The left panel is in units of the naive $\Delta$-pole contribution,
   $e^2/M_N^2 (M_\Delta-M_N)$, while the right panel is in physical units.
         }
\label{fig:beta}
\end{figure}
In nature, there is significant cancellation between the contribution 
from the $\Delta$-pole and from  chiral loops to the magnetic polarizability, a cancellation that is expected to diminish as the quark masses 
are increased~\protect\cite{Harald:2015CD} 
(the $\Delta$-nucleon mass splitting is relatively insensitive to the light-quark mass).
A large isovector component to the nucleon magnetic polarizability is found, and 
as  mentioned previously, while the isoscalar polarizabilities are subject to modifications (that are expected to be small) due to the omission of 
disconnected diagrams,  the isovector contributions are complete at the flavor-SU(3) symmetry point.  
It is interesting that, as shown in Fig.~\ref{fig:beta},
the magnetic polarizabilities of the light nuclei are found to be  near that of the proton.

The precision with which we have been able to determine the neutron systems is sufficient to determine that, while the di-neutron is bound at these heavier quark masses, there are values of the magnetic field for which it unbinds and the ground state becomes two isolated neutrons.
This is a QCD Feshbach resonance!
\begin{figure}[!ht]
  \centering  
  \includegraphics[width=0.6\columnwidth]{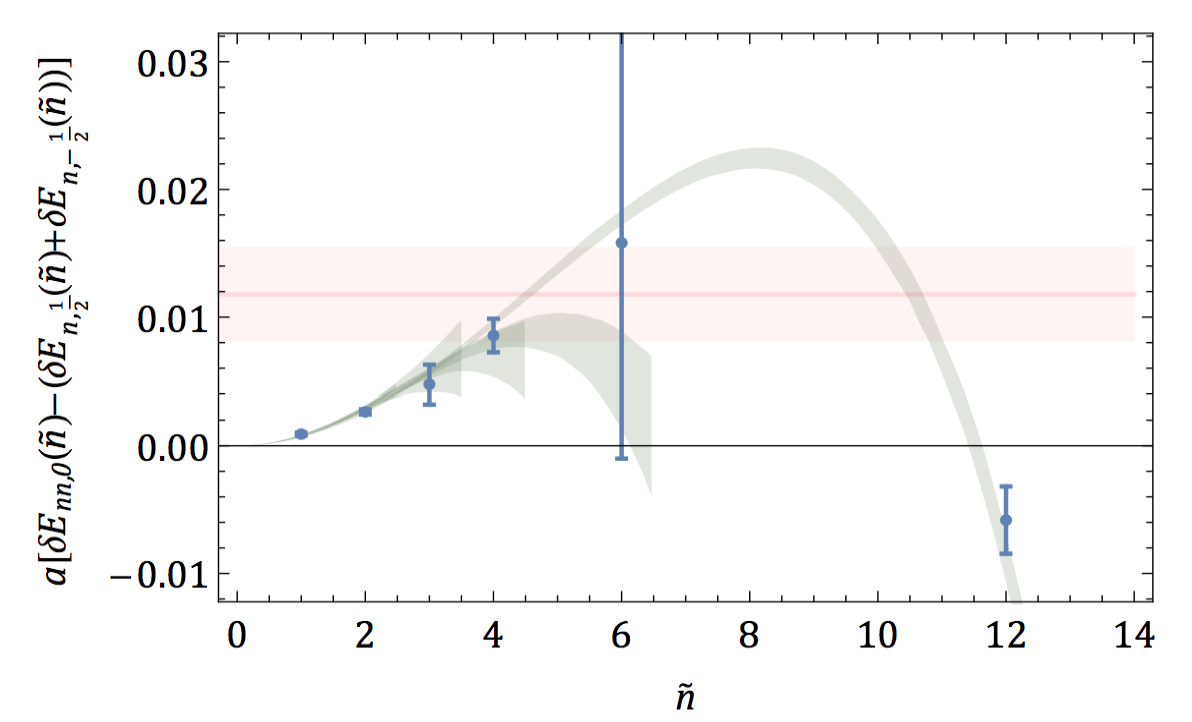}
 \vskip -0.15in
  \caption{
  The   energy difference (in lattice units) between two neutrons and  the bound di-neutron as a function of magnetic field 
  strength~\protect\cite{Chang:2015qxa}.   
  The red band corresponds to breakup threshold.  }
\label{fig:nnF}
\end{figure}
Figure~\ref{fig:nnF} shows the results of the LQCD calculation at $m_\pi=805~{\rm MeV}$ of the energy difference between the bound di-neutron and two isolated neutrons.  
Clearly, the di-neutron is becoming less bound with increasing magnetic field, 
and is consistent with being unbound, and hence the two-neutron system having an infinite scattering length, near $\tilde n\sim 5$.

In summary, Lattice QCD is emerging from an extended research and development phase into the production phase.  
Calculations of the binding and properties of light nuclei are now possible, and I have presented the state-of-the-art of such calculations.
The magnetic moments are providing important insights into the nature of nuclei and their stability with regard to changes in the 
fundamental parameters of nature.
The first inelastic nuclear reaction, $np\rightarrow d\gamma$, has been calculated and when extrapolated to the physical point is found to be in agreement 
with experiment. 
The magnetic polarizabilities of light nuclei have been calculated, and a large isovector component has been identified in the nucleon.
These works
are merely a peek into the future
 precision era of QCD calculations of low-energy nuclear physics observables.

\vskip 0.1in
{\it
Calculations were performed using computational resources provided by the Extreme Science and Engineering Discovery Environment (XSEDE), which is supported by National Science Foundation grant number OCI-1053575, NERSC (supported by U.S. Department of Energy Grant Number DE-AC02-05CH11231), and by the USQCD collaboration. 
This research used resources of the Oak Ridge Leadership Computing Facility at the Oak Ridge National Laboratory, which is supported by the Office of Science of the U.S. Department of Energy under Contract No. DE-AC05-00OR22725. 
The PRACE Research Infrastructure resources Curie based in France at the Tr`es Grand Centre de Calcul 
and MareNostrum-III based in Spain at the Barcelona Supercomputing Center were also used. 
Parts of the calculations used the Chroma software suite~\cite{Edwards:2004sx}.  
This work was supported in part by DOE grant No. DE-FG02-00ER41132.
}

%%%%%%%%%%%%%%%%%%%%%%%%%%%%

\end{document}